# Never forget, whom was my ancestors: A cross-cultural analysis from Yonsei (fourth-generation Nikkei) in four societies using Data Mining

Alberto Ochoa-Zezzatti[1], Luana Hatsukimi[2], Hitomi Karuda[3], Julio Arreola[1] & Sandra Bustillos[1]

*Juarez City University[1], UNICAMP[2], Hokkaido University[3]*

*México[1], Brazil[2], Japan[3]*

This research explains the importance of transculturality in social networking in a wide variety of activities of our daily life. We focus our analysis to online activities that use social richness, analyzing societies in Yakutia (A Russian Republic), Macau in China, Uberlândia in Brazil and Juarez City in Mexico, all with people descending from Japanese people. To this end, we performed surveys to gathering information about salient aspects of modernization and combined them using social data mining techniques to profile a number of behavioural patterns and choices that describe social networking behaviours in these societies.

We will define the terms "transculturality" and "multiculturality" as well as their contrast and roles in modern societies. Then we will describe a transculturality model that captures salient variables of modernization, and how these variables give raise to intervening aspects that end up shaping behavioural patterns in social networking. We will describe the data mining methodologies we used to extract these variables in each of these four societies including a survey conducted in these four countries, and provide a comparative analysis of the results in light of the proposed transculturality model.

We start by explaining the concept of transculturality and how this phenomenon affects, in different ways, people of differnet societies and culturally heterogenous. Furthermore, we explain the way of generating a correct analysis of these behaviours in social networkings to diferent activities of the daily life such as relationships with friends on line. Section 2 presents a transculturality model which explains the effects in people who exhibit this behavior. Section 3 presents a comparative analysis of samples of four different societies, as described in Section 4. In Section 5 we explain the results obtained with this analysis, and we conclude in Section 7 with conclusions about each topic analyzed.

I. Social Transculturality

Transcultural children of Third Culture (abbreviated TCKs or 3CKs for "third culture kids") who sometimes are also called "nomadic global," more culture talks about to that, like a boy, it has spent a period considerable of time in one or (s) which she is not the own one, reason why the integration of elements of those cultures and the own culture birth, within one third culture". The term TCK was first introduced by the sociologist Ruth Hill Useem in the 60's, and since then TCK have enjoyed the attention of researchers around the world. TCK tend to relate independent of their nationality. The composition of "the sponsors of America" (that is, the organization). Ever since the term was coined by the sociologist Ruth Hill Useem in the decade of 1960, TCK have become a world-wide subculture very well studied on different scieces as Social Psichology or Sociology of Minorities. TCK tend to have commonalities with others, independent of their nationality, which does with the nonTranscultural Children of their own country. The composition of the sponsors of America (that is to say, the organization whom it abroad sends to the family) changed much after World War II. Before World War II, 66% of of Transcultural Children came from families of the missionaries and 16% came from families of businesses. After World War II, with the increase of the international trade and the sprouting of two International superpowers, the composition of the international families has changed. To the sponsors usually it is divided in five categories: Missionary (17%), Businesses (16%), Government (23%), the Military (30%), and " Others" (14%) as children from Researcher arrivng to foreign Universities.

I.2 Multiculturality.

The concept of multiculturality is surprisingly similar to the concept of interculturality. It takes up the problems which different cultures have living together *within one society*. But there with the concept basically remains in the duct of the traditional understanding of culture; it proceeds from the existence of clearly distinguished, in themselves homogenous cultures - the only difference now being that these differences exist within one and the same state community, for example India or China with small minorities of less than 80 million people (Qinghai in China or Dadra & Nagar Haveli in India).

The concept seeks opportunities for tolerance and understanding, and for avoidance or handling of conflict. This is just as laudable as endeavors towards interculturality - but equally inefficient, too, since from the basis of the traditional comprehension of cultures a mutual understanding or a transgression of separating barriers cannot be achieved. As daily experience shows, the concept of multiculturality accepts and even furthers such barriers. Compared to traditional calls for cultural homogeneity the concept is progressive, but it is all too traditional understanding of cultures threatens to engender regressive tendencies which by appealing to a particularistic cultural identity lead to ghettoization or cultural fundamentalism.

II. Transculturality in Social Networkings.

**1. Macro-level: the altered cut of today's cultures**

a. Transculturality is, in the first place, a consequence of the *inner differentiation and complexity of modern cultures*. These encompass by development of diverse behavioral changes. These behaviors affect diverse changes in the day to day, furthermore, it is necessary to build a number of ways of life and cultures, which also interpenetrate or emerge from one another.

b. The old homogenizing and separatist idea of cultures has furthermore been surpassed through *cultures' external networking*. Cultures today are extremely interconnected and entangled with each other. Lifestyles no longer end at the borders of national cultures, but go beyond these, are found in the same way in other cultures. The way of life for an economist, an academic or a journalist is no longer German or French, but rather European or global in tone. The new forms of entanglement are a consequence of migratory processes, as well as of worldwide material and immaterial communications systems and economic interdependencies and dependencies. It is here, of course, that questions of power come in.

Consequently, the same basic problems and states of consciousness today appear in cultures once considered to be fundamentally different - think, for example, of human rights debates, feminist movements or of ecological awareness which are powerful active factors across the board culturally.

c. Cultures today are in general characterized by *hybridization*. For *every* culture, all *other* cultures have tendencially come to be inner-content or satellites. This applies on the levels of population, merchandise and information. Worldwide, in most countries, live members of all other countries of this planet; and more and more, the same articles - as exotic as they may once have been - are becoming available the world over; finally the global networking of communications technology makes all kinds of information identically available from every point in space.

Henceforward there is no longer anything absolutely foreign. Everything is within reach. Accordingly, there is no longer anything exclusively `own' either. Authenticity has become folklore, it is ownness simulated for others - to whom the indigene himself belongs. To be sure, there is still regional-culture rhetoric, but it is largely stimulatory and aesthetic; in substance everything is transculturally determined. Today in a culture's internal relations - among its different ways of life - there exists as much foreignness as in its external relations with other cultures.

**2. Micro-level: transcultural formation of individuals**

Transculturality is gaining ground moreover not only on the macrocultural level, but also on the individual's micro-level. For most of us, multiple cultural connexions are decisive in terms of our cultural formation. We are cultural hybrids. Today's writers, for example, emphasize that they're shaped not by a single homeland, but by differing reference countries, by Russian, German, South and North American, Korean or Japanese literature. Their cultural formation is transcultural (think, for example, of Naipaul or Rushdie) - that of subsequent generations will be even more so.

Sociologists have been telling us since the seventies that modern lives are to be understood "as a migration through different social worlds and as the successive realization of a number of possible identities" (Berger, Berger and Kellner, 1973: 77), and that we all possess "multiple attachments and identities" - "cross-cutting identities", as Bell put it (Bell, 1980: 243). What once may have applied only to outstanding persons like Montaigne, Novalis, Whitman, Rimbaud or Nietzsche, seems to be becoming the structure of almost everybody today.

III.- Proposal Model of Transculturality

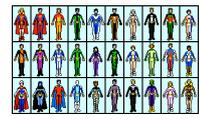

Salient aspects of Modernization

Intervening Variables

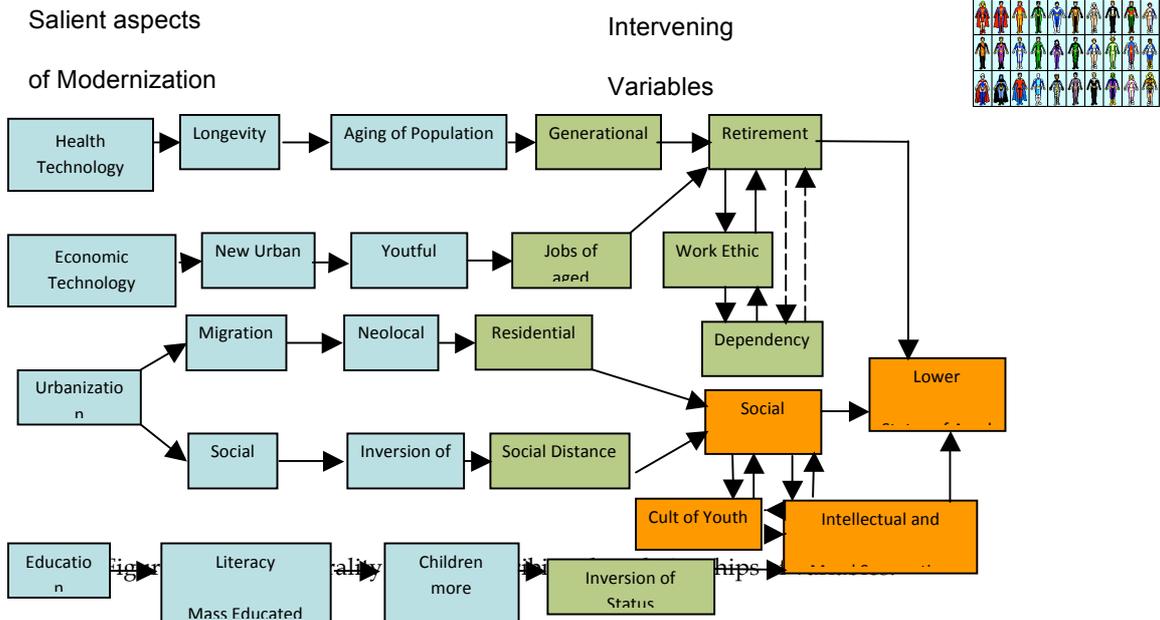

The proposal model of Transculturality is organized in three parts, first the most salient aspect of Modernization (in blue) denote the variables related with the actual situation in the life human, second the intervening variables (in green) determine the magnitude of affectation related with changes in social aspects and finally the resultant variables (in orange) which are a summary of the importance of Transculturality Behavior and determinar the effects of this effect in the future.

Transculturality and society are influenced by technology and culture, combining certain kind of co-dependency and culture simultaneously as cutting edge technology continues into modern world, printing press and computers, bringing around the so – called sciences of technology applied to economy and health issues.

In the modern world, superior technologies, resources, geography, and history give rise to robust economies; and in a well-functioning, robust economy, economic excess naturally flows into greater use of technology. Moreover, because technology is such an inseparable part of human society, especially in its economic aspects, funding sources for (new) technological endeavors are virtually illimitable. However, while in the beginning, technological investment involved little more than the time, efforts, and skills of one or a few men, today, such investment may involve the collective labor and skills of many millions.

Finally, other factors that influence the transcultural culture which are somehow related to educations are the institutions through the legislative factors like legislation, banning child marriages, the impact of western civilization and cultural diffusion, contact with people of different countries, the level of education and literacy in schools, the modernization of society by itself, the new attitudes to wealth, saving and risk taking, etc. Thus it is not possible to reach for social change and transculturality without education. Education has to do with beliefs and customs behind the transculturality issues which is adequate according to social progress. Take for example, the Hindu Code Bill Act, which remains due to the illiteracy of a large number of Indian people from generation to another and so on. The social changes caused by the political upheaval, technological progress and religious demand also education and a certain kind of social equilibrium.

IV.- A Comparative Analysis of Transculturality

Four different samples, all with les of 20 people respectively was analyzed to determine the similarities and differences in behaviors associated collective youth whom for some aspect of modernity, could be considered as a feature TCK because ethnic, religious, and migratory movements.

The transcultural factor of an individual is defined as:

$$v_{ij} = \alpha v_{ji,-1} + \beta_1 q_j^i + \sum_k \beta_k x_{jk} + \sum_l \gamma_l z_{il} + u_{ij}$$

(1)

In Where:
i= Culture expected
j= Culture received
  = Attributes associated with salient aspects of modernization
  = Attributes associated with Intervening Variables
  = Cost / benefit of Social Distance
k = Quality of life (depending on the given measurement rate)
l = Investment Status
  = Transculturality generation time
Z= Grade Technological Studies and the Individual Index

In Equation 1, we present a comparatva than in a society can be expected as a culture characterized largely received by each of the daily activities that determines a social group with a set of attributes associated with aspects of modernization sobresalietes a society and how this varies according to the externalities of the society in which it operates. The main experiment will be to detail each of the 150 issues (individuals), which will be represented by a Multiagent System, and a stop condition of 50 times (spaces in time that can vary according to a paradigm shift), this allow us to generate different scenarios Transculturality grade stock, it may be obtained after comparing the various similarities between the three companies, and determine the relationships between psychosocial each, finally using the similarity function shown in equation 2, generate the clusters associated with these features based on the concept of "Culturality".

$$\frac{\sum_{i=1}^{n} w_i \times sim(f_i^I, f_i^R)}{\sum_{i=1}^{n} w_i} \qquad (2)$$

In where:
$w_i$ determined for the importance weight of each attribute.
sim is the similarity function.
$f_i^I$ y $f_i^R$ son los valores del atributo i en el cluster de entrada (I) y en el cluster recuperado (R).

### V.- Results of our Comparative Analysis

The four samples representing different societies (Sample 1: Ciudad Juárez, a largest city in Mexico; Sample 2: Macau; Sample 3: Uberlândia, a largest city in Brazil and Sample 4: Yakutia, a largest Republic in Russian Federation) which was resuming using the next table:

**Table 1.** Differences on Transculturality behavior by gender for the four societies studied

|   | Sample 1 | | Sample 2 | | Sample 3 | | Sample 4 | |
|---|---|---|---|---|---|---|---|---|
|   | F | M | F | M | F | M | F | M |
| N | 5 | 7 | 8 | 7 | 9 | 6 | 4 | 3 |
| Health Technology | 22% | 22% | 21% | 26% | 29% | 27% | 21% | 37% |
| Economic Technology | 34% | 31% | 39% | 21% | 38% | 37% | 43% | 28% |
| Urbanization | 84% | 87% | 91% | 80% | 100% | 96% | 57% | 63% |
| Education | 25% | 50% | 43% | 63% | 38% | 66% | 63% | 58% |
| Longevity | 39% | 54% | 4% | 56% | 29% | 59% | 72% | 69% |
| New Urban Occupation | 78% | 85% | 75% | 81% | 95% | 91% | 55% | 57% |
| Migration | 83% | 93% | 82% | 98% | 76% | 93% | 63% | 68% |
| Social Mobility | 69% | 91% | 71% | 79% | 71% | 87% | 73% | 64% |
| Literacy | 31% | 48% | 29% | 56% | 48% | 71% | 82% | 89% |
| Mass Educated | 72% | 89% | 89% | 91% | 81% | 87% | 89% | 94% |
| Technology Trends | 19% | 57% | 32% | 40% | 24% | 26% | 58% | 32% |
| Aging of Population | 83% | 67% | 86% | 74% | 76% | 70% | 84% | 81% |
| Youtful Aspects | 81% | 77% | 89% | 70% | 76% | 70% | 72% | 76% |
| Neolocal Marriage | 61% | 87% | 75% | 86% | 71% | 80% | 68% | 73% |
| Inversion of Status | 71% | 74% | 100% | 100% | 100% | 100% | 83% | 81% |
| Children more educated than parents | 11% | 17% | 14% | 44% | 19% | 60% | 45% | 57% |
| Generational competition | 80% | 72% | 71% | 86% | 71% | 91% | 84% | 89% |
| Jobs of Aged obsoleted | 45% | 47% | 43% | 72% | 29% | 79% | 64% | 61% |
| Residential segregation | 88% | 93% | 100% | 98% | 90% | 100% | 74% | 69% |
| Social Distance | 48% | 45% | 75% | 88% | 52% | 80% | 68% | 75% |
| Inversion of Status | 27% | 33% | 39% | 65% | 100% | 62% | 74% | 71% |
| Retirement | 35% | 29% | 18% | 44% | 33% | 59% | 48% | 46% |
| Work Ethic | 19% | 14% | 11% | 16% | 19% | 21% | 35% | 29% |
| Dependency | 44% | 36% | 22% | 27% | 18% | 16.5% | 23% | 28% |
| Social Segregation | 21% | 24% | 45% | 49% | 37% | 39% | 52% | 44% |
| Cultural of Youth | 35% | 47% | 67% | 73% | 62% | 69% | 57% | 73% |
| Lower Status of Aged | 55% | 53% | 44% | 38% | 32% | 37% | 38% | 43% |
| Intelectual and Moral Segregation | 27% | 31% | 55% | 51% | 67% | 58% | 63% | 72% |
| *Sum of Transculturality (mean)* | 5.6 | 7.2 | 12 | 14 | 11.1 | 14.9 | 12.7 | 10.3 |

These values describe the relationship of different attributes related to the transculturality, by sample. The most relevant are:

1. The sample 1 from Juarez City is characterized by the intention to change stablished paradigms related to the social position – in this case people born in Mexico but with continuosly travels to Japan, in many cases descending from immigrants from different societies in the south and central states to Mexico is an adventage in social circles as Social Clubs and Foundations, the lastnames with Japanese roots has the perspective of fascination in border society. Only three persons of this sample descending from people only from Juarez City, a situation commonly on this border city,

The sample 2 from Macau in China, has a variety of people with ancestors arriving principaly from Asian and a few ancestors from Japan and Portugal, many of these individuals miss the ancestral costumes and try to organize events to remain cultural closely, this sample confronts the ideologic perspective, because Spain has different languages established officialy. In the case of social networkings, these individuals has an extended social networking with many different persons, try to be tolerants with them and respect the majority of different ideas, in recent years the diverses Natioalisms cofront at new generations to think really about the independence from China as in the case of Hong Kong.

The sample 3 from Uberlândia in Brazil, has a variety of people with ancestors arriving principaly from diverse socities than different and culturally variety as Portugal, German, Italy and Lebanon inclusive Palestine and with very strong roots about the family, this sample is featured with a strong relationship with their culture, religion and language. In specially with the traditional food, in the colective imaginary any person descending from this kind of people need cooking this styles of food to specially festivals and celebrations during all year.

The sample 4 from Yakutia, has people with ancestors from diverse places, people with origins in Asian specially Korean, Eastern Europe, and Russian including these group. The majority of these persons have relationship with their traditional heritage through of food, clothes, language and participation of Collectives which reflect the importance of be proud about their different ideas and point of views related with the life style. In the case of social networkings, they use different language to express information to their pals and using many time photos or albums to describe travels or hobbies.

V.- Results of our Comparative Analysis

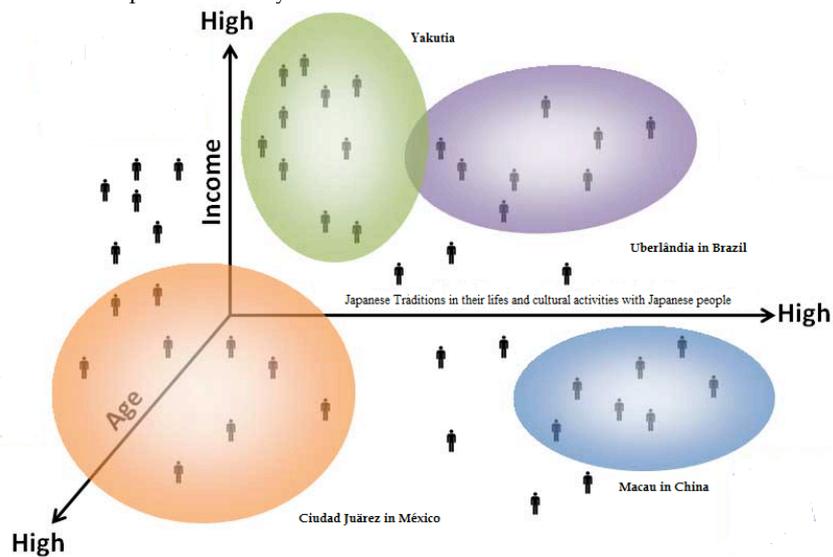

Figure 2.- Clusters associated at status of Transculturality including the four samples.

Using the location of each cluster (the four samples generated a variety of Clusters) is possible understand the variety of kind of transculturality reflected in many diary activities, the blue cluster conformed with Macau people, violet composed with Uberlândia people in Brazil, orange including Juarez City people –a society in Mexico-, and green integrated with Yakutia people, all these clusters and theirs variations respectively grouping the majority of young people disgusts to leave old traditional costumes of their family including change their names to more modern and Western style, see Figure 2. The other clusters and theirs variations are characterized by a major age and learn Japanese and traditional costumes with their family when they were lad or lass.

Using different colors which represented at HDI (Human Developed Index), we generate a lot of clusters which organize the totally of individuals, in the upper corner at left the more transculturally individuals, and the low corner at right the individuals with less transculturality behavior. The location and the color size the magnitude and intensity of Transculturality Behavior.

VI.- Profiles generated by the previous analysis

Descendents from Japanese people, is a society with specialized skills with modern characteristics would therefore a scarlet ease to be able to adapt to external influences, for it is they should know more than one language, one international trip a year, meet different people from different societies to which they live. The big difference between Multiculturality and Transculturality are determined by the ability to integrate possessing ie how an individual feels "socially cohesive" with the rest of the individuals in these societies, minorities bear a great "symbolic capital "because they are a unique people in more than one case is seen as a bulwark socio-cutural, when the sample was selected was chosen on the Republic of Yakutia, since that is society has developed in tandem, nationalist sentiment use of Yakutsk and cultural practices brought by immigrants from Yakutia. Multivariable analysis the effect of educational institution in private and public sectors based on the application of multilevel of regression techniques on data from OECD PISA-2011, which supports the working hypothesis that educational institutions play a key role in the transculturality phenomena. Generally speaking, Multivariable methodologies uses the development of data collection instruments, test translations, sample design, data analysis, and reporting the results. PISA studies represents the first incursion of OECD in a complex world of internatioal evaluations since 1964. Those can be considered as an educational lab worlwide and they basically touch on individuals performance based on the transcultirality among those, by taking into accounto the economic and cultural issues.

VII.- Conclusions and Future Research

Describe a subjective concept as Transculturality is very complicated, because exist a plethora of activities related with this social concept. An individula will be recognized in another society, for example Diego Urbina whom is the first Italian Cosmonaut borned and lived many time in Colombia with his Father family, but when arrive in Italy, the Italian Mother family projected at him to reach many skills to be the first Italian Cosmonaut to emulate a travel to Mars. The majority of activities related with Transculturality take social advantages, for example the use of English language is an advantage in the laboral market and the status of American ciudadany is a very important achievement which generate a status in a border society, this implies to confront with a few problems in relation with another person.

In addition, data mining techniques will be developed to obtain profiles and patterns of sexual offenders while operating inside the virtual worlds. We consider that effective medium and long term measures to counter illegal activities of sexual offenders in virtual worlds include the development of artificial intelligence programs combined with human searches and analyses. Ultimately, humans will have the last word and will make the final decision about what course of action to take to prevent and counter sexual offenders in virtual worlds, supported by data analyses generated by artificial intelligence programs.